%% file: eDDMPC_conf.tex
\documentclass[conference]{ieeeconf}
\usepackage{algorithm,algorithmic}
\usepackage[noadjust]{cite}
\usepackage{amssymb,mathrsfs, mathtools, float}

\DeclarePairedDelimiter{\norm}{\lVert}{\rVert}

\newtheorem{theorem}{Theorem}
\newtheorem{lemma}{Lemma}

\newtheorem{remark}{Remark}

\newtheorem{corollary}{Corollary}
\newtheorem{definition}{Definition}
\usepackage{tikz}
\usetikzlibrary{matrix,decorations.pathreplacing,calc}
\IEEEoverridecommandlockouts

\makeatletter
\let\NAT@parse\undefined
\newcommand{\removelatexerror}{\let\@latex@error\@gobble}
\makeatother
\usepackage{hyperref}

\newcommand\copyrighttext{%
	\footnotesize \copyright 2024 IEEE. Personal use of this material is permitted. Permission from IEEE must be obtained for all other uses, in any current or future media, including reprinting/republishing this material for advertising or promotional purposes, creating new collective works, for resale or redistribution to servers or lists, or reuse of any copyrighted component of this work in other works.}
\newcommand\copyrightnotice{%
	\begin{tikzpicture}[remember picture,overlay]
		\node[anchor=south,yshift=5pt] at (current page.south) {\fbox{\parbox{\dimexpr\textwidth-\fboxsep-\fboxrule\relax}{\copyrighttext}}};
	\end{tikzpicture}%
}

\begin{document}
		
\title{\LARGE{\textbf{Sample- and computationally efficient data-driven predictive control}}}

\author{Mohammad Alsalti$^*$, Manuel Barkey$^*$, Victor G. Lopez and Matthias A. Müller%
	\thanks{Leibniz University Hannover, Institute of Automatic Control, 30167 Hannover, Germany. E-mail: alsalti@irt.uni-hannover.de, barkey@irt.uni-hannover.de, lopez@irt.uni-hannover.de, mueller@irt.uni-hannover.de.%
	}
	\thanks{This work has received funding from the European Research Council (ERC) under the European Union’s Horizon 2020 research and innovation programme (grant agreement No 948679).
	}
	\thanks{$^*$Mohammad Alsalti and Manuel Barkey contributed equally to this work.
	}
}

	\maketitle%
	\thispagestyle{empty}%
	\copyrightnotice%
	\begin{abstract}
		Recently proposed data-driven predictive control schemes for LTI systems use non-parametric representations based on the image of a Hankel matrix of previously collected, persistently exciting, input-output data. Persistence of excitation necessitates that the data is sufficiently long and, hence, the computational complexity of the corresponding finite-horizon optimal control problem increases. In this paper, we propose an efficient data-driven predictive control (eDDPC) scheme which is both more sample efficient (requires less offline data) and computationally efficient (uses less decision variables) compared to existing schemes. This is done by leveraging an alternative data-based representation of the trajectories of LTI systems. We analytically and numerically compare the performance of this scheme to existing ones from the literature.
	\end{abstract}
	\IEEEpeerreviewmaketitle
	
	\input{sections/introduction}
	\input{sections/preliminaries}
	\input{sections/eDDPC}
	\input{sections/comparison}
	\input{sections/conclusions}
	
	\bibliographystyle{IEEEtran}
	\bibliography{references}
	
\end{document}

%% file: sections/introduction.tex
\section{Introduction}\label{sec_introduction}
Model predictive control (MPC) \cite{Rawlings17} is an optimization-based control technique that involves solving a finite-horizon optimal control problem repeatedly at each time instant, and then applying the first part of the optimal input to the system. In recent years, the development of data- and learning-based MPC schemes has received a significant amount of attention. Here, the unknown system model or the MPC controller itself are either approximated using machine learning techniques or directly computed using input-output data of the system, see, e.g. \cite{Hewing20, Markovsky21} and the references therein.

One prominent approach to data-driven system analysis and control exploits the behavioral approach to systems theory~\cite{Willems86}. For instance, Willems' fundamental lemma \cite{Willems05} provides a non-parametric representation of the finite-length behavior of a discrete-time LTI system as the image of a Hankel matrix of a single persistently exciting (PE) trajectory. In particular, it asserts that any trajectory of an LTI system can be expressed as a linear combination of time-shifts of a single measured trajectory. This result was successfully used for system analysis and control design for linear systems and extended to classes of nonlinear systems. The reader is referred to \cite{Markovsky21} and \cite{Martin23} for a comprehensive review. 

Among several successful applications of the fundamental lemma is its use in data-driven predictive control schemes (DDPC). Specifically, instead of relying on the typical steps of deriving and identifying a predictive model for the MPC, one can use previously collected input-output data as a data-based non-parametric predictive model. This was first proposed in \cite{Yang15} and further developed in \cite{Coulson20}. Theoretical guarantees were established in \cite{Berberich203, Berberich22_IRP}. Extensions of this data-driven predictive control scheme to classes of nonlinear systems appeared in \cite{Berberich204, Alsalti2021c, Verhoek21, Lian21}. Several other extensions and applications of this scheme include: stochastic DDPC \cite{Pan22, Breschi23}, segmented DDPC \cite{Dwyer23}, distributed DDPC \cite{Kohler22, Allibhoy21} and generalized DDPC \cite{Lazar23}, among many others.

Crucial to the successful application of this scheme is that the offline collected data is persistently exciting (PE). This, however, necessitates that the data is sufficiently long, and this required number of data points increases with increased system order, number of inputs and prediction horizon length. As a result, the computational complexity of solving the corresponding finite-horizon optimal control problem increases. Moreover, it may not be easy to obtain such PE data in practice. For instance, the length of the available data might not be long enough to allow for application of DDPC schemes with long prediction horizons. In such circumstances, it was shown in \cite{Alsalti2023_md} how one can obtain alternative non-parametric representations of LTI systems using a small number of (potentially irregularly measured) data points. In this work, we exploit the proposed representation in \cite{Alsalti2023_md} and illustrate its use in an efficient DDPC scheme.

The contributions of this paper are as follows: we propose an efficient data-driven predictive control (eDDPC) scheme which is both sample efficient (requires less offline data) and computationally efficient (includes less decision variables) compared to existing schemes. We show that the closed-loop behavior is equivalent to that of existing schemes and hence enjoys the same theoretical guarantees. We then analytically and numerically compare the performance of the proposed scheme with other existing ones from the literature and illustrate how it consistently outperforms them.

The paper is structured as follows: Section~\ref{sec_preliminaries} introduces notation and necessary preliminaries. In Section~\ref{sec_eDDPC}, we formulate the eDDPC scheme. Section~\ref{sec_eDDPC_comparison} includes analytical and numerical comparisons to existing schemes and Section~\ref{sec_conclusions} concludes the paper.

%% file: sections/preliminaries.tex
\section{Preliminaries}\label{sec_preliminaries}
The sets of integers, natural and real numbers are denoted by $\mathbb{Z},\mathbb{N},\mathbb{R}$, respectively. The restriction of integers to an interval is denoted by $\mathbb{Z}_{[a,b]}$, for $b>a\in\mathbb{Z}$. We use $I_m$ to denote an $m\times m$ identity matrix and $0_{n\times m}$ to denote an $n\times m$ matrix of zeros. When the dimensions are clear from the context, we omit the subscript for simplicity. For a matrix $M\in\mathbb{R}^{m\times n}$, we denote its image by im$(M)$ and its kernel by ker$(M)$. When a basis of ker$(M)$ is to be computed, we write $N=~\textup{null}(M)$ which returns a matrix $N$ of appropriate dimensions such that $MN=~0$. For a symmetric positive definite matrix $P=P^\top\succ0$, we define the weighted norm of a vector $x$ as $\norm*{x}_P\coloneqq\sqrt{x^\top Px}$. In contrast, $\norm*{x}_i$, $i\in\{1,2,\infty\}$, denote the standard $\ell_1,\ell_2$ (Euclidean), and $\ell_\infty$ norms, respectively.

The set of infinite length, $q-$variate, real-valued time series $w=(w_0,w_1,\ldots)$ is denoted by $\left(\mathbb{R}^{q}\right)^\mathbb{N}$. For $T\in\mathbb{N}$, the set of finite-length, $q-$variate, real-valued time series $w=(w_0,w_1,\ldots,w_{T-1})$ is denoted by $\left(\mathbb{R}^{q}\right)^T$.
With slight abuse of notation, we also use $w$ to denote the stacked vector $w=\begin{bmatrix}w_0^\top & w_1^\top & \cdots & w_{T-1}^\top\end{bmatrix}^\top\in~\mathbb{R}^{qT}$, and a window of it by $w_{[a,b]}$ where $0\leq a < b \leq T-1$. The Hankel matrix of depth $L$ of $w$ is defined as
\begin{equation*}
	\mathscr{H}_L(w) = \begin{bmatrix} w_{[0,L-1]} & w_{[1,L]} & \cdots & w_{[T-L,T-1]}
	\end{bmatrix}.\label{eqn_Hankelmatrix}
\end{equation*}

The (finite-length) behavior ($\mathscr{B}|_T$) $\mathscr{B}$ of a dynamical system is defined as the set of all (finite-) infinite-length trajectories that can be generated by the system (cf. \cite{Willems86}). A trajectory of length $T$ of the system is denoted by $w\in\left.\mathscr{B}\right|_T\subset\left(\mathbb{R}^q\right)^{T}$. The system is linear if $\mathscr{B}$ is a subspace and it is time-invariant if it is invariant to the action of the \textit{shift operator}, defined as $\sigma^j w(k) \coloneqq w(k+j)$ for $j\in\mathbb{N}$.

Let $u_t\in\mathbb{R}^m$ and $y_t\in\mathbb{R}^p$ denote the inputs and outputs of system $\mathscr{B}$ at time $t$. We define a partitioning of the variable $w_t\in\mathbb{R}^q$ such that $w_t=\begin{bsmallmatrix} u_t \\ y_t \end{bsmallmatrix}$. The set of discrete-time LTI systems with $q$ variables and bounded complexity $(m,n,\ell)$ is denoted by $\mathscr{L}_{m,n,\ell}^{q}$, where $q=m+p$ and $(m,n,\ell)$ denote (i) the number of inputs $\textbf{m}(\mathscr{B}) \leq m$, (ii) the order of the system $\textbf{n}(\mathscr{B}) \leq n$, and (iii) the lag of the system $\boldsymbol{\ell}(\mathscr{B}) \leq \ell$, which is the observability index in the state-space framework. These integers satisfy the following relation $\ell\leq n\leq p\ell$ \cite{Markovsky22}. 

A finite-dimensional LTI system $\mathscr{B}\in\mathscr{L}_{m,n,\ell}^{q}$ admits a kernel representation \cite[Part I]{Willems86}
\begin{equation}
	\mathscr{B} = \textup{ ker } R(\sigma) \coloneqq \{ w:\mathbb{Z}_{\geq0}\to\mathbb{R}^q ~|~ R(\sigma)w=0\},
	\label{kernel_rep}
\end{equation}
where the operator $R(\sigma)$ is defined by the polynomial matrix
\begin{align}
	R(z) &= \begin{bmatrix}
		r_1(z)\\
		\vdots\\
		r_g(z)
	\end{bmatrix}\hspace{-1mm} = \hspace{-1mm}\begin{bmatrix}
		r_{1,0} + r_{1,1}z + \ldots + r_{1,\ell_1}z^{\ell_1}\\
		\vdots\\
		r_{g,0} + r_{g,1}z + \ldots + r_{g,\ell_g}z^{\ell_g}
	\end{bmatrix},\label{eqn_kernelpolymatrix}
\end{align}
with $r_{i,j}\in\mathbb{R}^{1\times q}$. Given a trajectory $w\in\mathscr{B}|_T$ of an LTI system $\mathscr{B}\in\mathscr{L}_{m,n,\ell}^{q}$, it holds by linearity and shift-invariance that im$(\mathscr{H}_L(w)) \subseteq \mathscr{B}|_L$. When equality holds, we obtain a data-based representation of the finite-length behavior of the system. For the case of controllable systems, Willems' fundamental lemma \cite{Willems05} provides conditions on the input, such that im$(\mathscr{H}_L(w)) = \mathscr{B}|_L$. This condition is known as persistence of excitation (PE) and is defined as follows.
\begin{definition}\label{def_PE}
	A sequence $u\in(\mathbb{R}^m)^{T}$ is said to be persistently exciting of order $L$ if $\textup{rank}(\mathscr{H}_L(u))=mL$.
\end{definition}
\begin{lemma}\textup{\cite{Willems05}}\label{lemma_FL}
	Let $w\in\mathscr{B}|_T$ with $\mathscr{B}\in\mathscr{L}_{m,n,\ell}^{q}$ controllable. Let $u\in(\mathbb{R}^m)^T$ be PE of order $L+n$, then $\textup{rank}(\mathscr{H}_L(w)) = mL+n$, and $\bar{w}\in\mathscr{B}|_L$ if and only if there exists $\alpha\in\mathbb{R}^{T-L+1}$ such that
		\begin{equation}
		\mathscr{H}_L(w)\alpha = \bar{w}.\label{GPE_FL}
	\end{equation}
\end{lemma}
This lemma states that each length-$L$ trajectory of a controllable LTI system can be represented as a linear combination of time-shifts of previously collected data $w$ (which correspond to the columns of the Hankel matrix $\mathscr{H}_L(w)$), given that the input is sufficiently exciting. We now recall the following result, which allows us to retrieve a kernel representation \eqref{eqn_kernelpolymatrix} of an LTI system directly from data.
\begin{corollary}\textup{\cite[Cor. 2]{Alsalti2023_md}}\label{cor_kernelRd}
	Let $w\in\mathscr{B}|_T$ where $\mathscr{B}\in\mathscr{L}_{m,n,\ell}^{q}$. Let $d\geq\ell+1$, and suppose $\textup{rank}(\mathscr{H}_d(w))=md+n$. Then, the coefficients of the polynomial matrix $R(\sigma)$ in \eqref{eqn_kernelpolymatrix} are given by the rows of $R_d\in\mathbb{R}^{pd-n\times qd}$ which satisfies $R_d\mathscr{H}_d(w)=0$.
\end{corollary}
By exploiting the kernel structure of Hankel matrices, it was shown in \cite{Alsalti2023_md} that one can obtain a full column rank matrix whose image is equal to the finite-length behavior of the system $\mathscr{B}|_L$. This result is summarized in the following theorem, and a sketch of the proof is provided for completeness.
\begin{theorem}\textup{\cite{Alsalti2023_md}}\label{thm_AFL}
	Given $w\in\mathscr{B}|_T$ where $\mathscr{B}\in\mathscr{L}_{m,n,\ell}^{q}$ is controllable, let $u\in(\mathbb{R}^m)^T$ be persistently exciting of order $d+n$ for $d=\ell+1$. Then, for any $L\geq d$, $\bar{w}\in\mathscr{B}|_L$ if and only if there exists a vector $\beta\in\mathbb{R}^{mL+n}$ such that
	\begin{equation}
		P\beta=\bar{w},\label{eqn_efficientFL}
	\end{equation}
	where $P=\textup{null}(\Gamma)$ and $\Gamma$ is given by
	\begin{align}
		\hspace{-1mm}\Gamma \hspace{-1mm}=\hspace{-2mm}\label{Bmatrix}\begin{tikzpicture}[decoration={brace,amplitude=5pt},baseline=(current bounding box.west), scale=0.95, every node/.style={scale=0.9}]
			\node[align=center] at (0,0) {$\begingroup\setlength\arraycolsep{2.5pt}\begin{bmatrix}
					\begin{matrix}
						r_{1,0}\\[-0.25em]
						r_{2,0}\\[-0.25em]
						\vdots\\[-0.25em]
						r_{pd-n,0}
					\end{matrix} & \begin{matrix}
						r_{1,1}\\[-0.25em]
						r_{2,1}\\[-0.25em]
						\vdots\\[-0.25em]
						r_{pd-n,1}
					\end{matrix} & \begin{matrix}
						\cdots\\[-0.25em] \cdots\\[-0.25em] \ddots\\[-0.25em] \cdots
					\end{matrix} & \begin{matrix}
						r_{1,d-1}\\[-0.25em]
						r_{2,d-1}\\[-0.25em]
						\vdots\\[-0.25em]
						r_{pd-n,d-1}
					\end{matrix} &\\[-0.25em]
					& \begin{matrix}
						r_{1,0}\\[-0.25em]
						r_{2,0}\\[-0.25em]
						\vdots\\[-0.25em]
						r_{p,0}
					\end{matrix} & \begin{matrix}
						r_{1,1}\\[-0.25em]
						r_{2,1}\\[-0.25em]
						\vdots\\[-0.25em]
						r_{p,1}
					\end{matrix} & \begin{matrix}
						\cdots\\[-0.25em] \cdots\\[-0.25em] \ddots\\[-0.25em] \cdots
					\end{matrix} & \begin{matrix}
						r_{1,d-1}\\[-0.25em]
						r_{2,d-1}\\[-0.25em]
						\vdots\\[-0.25em]
						r_{p,d-1}
					\end{matrix}\\[-0.25em]
					& &\ddots & \ddots & \ddots & \ddots\\[-0.25em]
					& & & \begin{matrix}
						r_{1,0}\\[-0.25em]
						r_{2,0}\\[-0.25em]
						\vdots\\[-0.25em]
						r_{p,0}
					\end{matrix} & \begin{matrix}
						r_{1,1}\\[-0.25em]
						r_{2,1}\\[-0.25em]
						\vdots\\[-0.25em]
						r_{p,1}
					\end{matrix} & \begin{matrix}
						\cdots\\[-0.25em] \cdots\\[-0.25em] \ddots\\[-0.25em] \cdots
					\end{matrix} & \begin{matrix}
						r_{1,d-1}\\[-0.25em]
						r_{2,d-1}\\[-0.25em]
						\vdots\\[-0.25em]
						r_{p,d-1}
					\end{matrix}
				\end{bmatrix}\endgroup$};
			\draw[decorate] (2,1.25) -- (3.5,-1) node[above=5pt,midway,sloped] {$L-d$ \textup{times}};
		\end{tikzpicture}\hspace{-1mm},
	\end{align}
	and $r_{i,j}\in\mathbb{R}^{1\times q}$ are the elements of the matrix $R_d$ in Cor.~\ref{cor_kernelRd}.
\end{theorem}
\begin{proof}
	Since the input is persistently exciting of order $d+n$, it holds by Lemma~\ref{lemma_FL} that rank$(\mathscr{H}_d(w))=md+n$. Therefore, by Corollary~\ref{cor_kernelRd}, one can compute a basis for the left kernel of this matrix (i.e., $R_d\mathscr{H}_d(w)=0$) which specifies a kernel representation of the system. The rest of the proof follows from \cite[Lemma 2]{Alsalti2023_md} and \cite[Theorem 3]{Alsalti2023_md}, where it was shown that im$(P)=\mathscr{B}|_L$, for any $L\geq d$, with $P$ defined as in the theorem statement.
\end{proof}
Theorem~\ref{thm_AFL} provides a data-based representation of the trajectories of an LTI system, but is different from the one provided by Lemma~\ref{lemma_FL}. This alternative representation has two advantages: (i) it can be obtained from a (very) small number of data points. In particular, we only need input data that is PE of order $d+n$ (independent of $L$). This is achieved by computing $R_d$ as the left kernel of $\mathscr{H}_d(w)$ as in Corollary~\ref{cor_kernelRd} (note that this can even be done if the data $w$ potentially contains irregularly missing values, cf. \cite{Alsalti2023_md}), which allows us to represent any trajectory $\bar{w}$ of length $L$ (for arbitrary $L\geq d$) via \eqref{eqn_efficientFL}. (ii) It does not result in an over-parameterization of the spanned input-output trajectories. Specifically, the regressor vector $\beta$ in \eqref{eqn_efficientFL} is of dimension $mL+n$ only, i.e., independent of the number of data points.

In Section~\ref{sec_eDDPC_nominal} we comment in more detail on the similarities and differences between Theorem~\ref{thm_AFL} and Lemma~\ref{lemma_FL}. Then, we explain how we make use of those differences to propose an efficient data-driven predictive control scheme.

%% file: sections/eDDPC.tex
\section{Efficient DDPC}\label{sec_eDDPC}
In this section, we propose an efficient data-driven predictive control scheme which is both sample efficient (uses lass offline data points) and computationally efficient (includes less decision variables) compared to existing schemes. This is done by using the alternative data-based representation of LTI systems in Theorem~\ref{thm_AFL} as a predictor in a receding horizon scheme. In Section~\ref{sec_eDDPC_overview}, we start by giving an overview of existing DDPC schemes and then in Section~\ref{sec_eDDPC_nominal} we introduce our proposed eDDPC scheme in the nominal (noise-free) case.
\subsection{Overview of existing schemes}\label{sec_eDDPC_overview}
As proposed in \cite{Yang15, Coulson20}, a data-driven predictive control scheme for LTI systems relies on the non-parametric representation in \eqref{GPE_FL} of the finite-length trajectories of an LTI system to make predictions over a finite horizon. Later in \cite{Berberich203}, terminal equality constraints were introduced  in order to exponentially stabilize a (known) equilibrium point of an unknown LTI system. Such an equilibrium point is defined in terms of the system's inputs and outputs as follows. 
\begin{definition}\cite{Berberich203}
	We say that $w^s$ is an equilibrium of $\mathscr{B}\in\mathscr{L}_{m,n,\ell}^{q}$ if the sequence $\bar w\in(\mathbb{R}^q)^{n+1}$ with $\bar w_k=w^s$, for all $k\in\mathbb{Z}_{[0,n]}$, is a trajectory of the system, i.e., $\bar{w}\in\mathscr{B}|_{n+1}$.
\end{definition}

The DDPC scheme in \cite{Berberich203} repeatedly solves, at each time step $t$, the following finite-horizon optimal control problem
\begin{equation}
	\begin{aligned}
		\min_{\alpha(t),\bar w(t)}\quad 
		& \sum_{k=0}^{L-1}\norm*{\bar w_k(t) - w^s}_W^2\\
		\textrm{s.t.}\quad 
		& \bar w_{[-n,L-1]}(t) = \mathscr{H}_{L+n}(w^{\textup{data}})\alpha(t)\\
		& \bar w_{[-n,-1]}(t) = w_{[t-n,t-1]}\\
		& \bar w_{[L-n,L-1]}(t) = w^s_n\\
		& \bar w_k(t) \in \mathbb{W},\quad \forall k \in \mathbb{Z}_{[0,L-1]}.
	\end{aligned}
	\label{eqn_DDPC}
\end{equation}
The notation in \eqref{eqn_DDPC} is summarized as follows: $w^{\textup{data}}\in(\mathbb{R}^q)^T$ refers to the a priori collected input-output data, while $\bar w(t)\in(\mathbb{R}^q)^L$ refers to the predicted input-output trajectories (over the horizon length $L$) at time $t$. The online measurements are denoted by $w_t$. The stage cost is a quadratic function that penalizes the deviation from the given set point (for some weighting matrix $W\succ0$). We use $w^s_n$ to denote an $n-$dimensional vector containing $n$ instances of $w^s$. Finally, $\mathbb{W}$ denotes the constraints set and is defined as follows
\begin{equation}
	\mathbb{W}\coloneqq\{w=\begin{bsmallmatrix}
		u\\y
	\end{bsmallmatrix}~|~ u\in\mathbb{U},\,y\in\mathbb{Y}\},
\end{equation}
where $\mathbb{U},\mathbb{Y}$ denote the input and output constraint sets, respectively, with $w^s\in\textup{int}(\mathbb{W})$. 

To implement the DDPC scheme, one requires an offline data sequence $w^{\textup{data}}\in(\mathbb{R}^q)^T$, where the input is PE of order $L+2n$ ($L\geq n$), as well as an initial trajectory $w_{[t-n,t-1]}$ to fix the internal state. Notice that, unlike Lemma~\ref{lemma_FL}, persistence of excitation here is required to be of order $L+2n$ instead of $L+n$. This is because the length of the predicted trajectories is extended by $n$ instances to account for the initialization step, i.e., $\bar w_{[-n,-1]}(t) = w_{[t-n,t-1]}$ in \eqref{eqn_DDPC}. Persistence of excitation necessitates that the Hankel matrix of the input $\mathscr{H}_{L+n}(u^{\textup{data}})$ has at least as many column as rows, and hence, $T\geq(m+1)(L+2n)-1$ (cf. Definition~\ref{def_PE}). Therefore, implementing a DDPC scheme with a large prediction horizon requires a longer sequence of offline collected data. However, in many practical settings, it may be difficult to collect such a long sequence, or one may only have a short sequence of data which is not enough to span trajectories of length $L+n$.

In \cite{Dwyer23}, the authors propose a partitioning of the (long) prediction horizon to multiple ($n_s$) segments of length $T_{\textup{ini}}\geq \ell$. If the available data is long enough to span trajectories of length $T_{\textup{ini}}$, then one can restructure the original DDPC (with longer horizon) to obtain the complete predicted trajectories. This partitioning increases the number of decision variables in the optimization problem at each time $t$, and the segmented scheme only shows better computational performance than standard DDPC when the prediction horizon is significantly large. In contrast, our proposed scheme (see Section~\ref{sec_eDDPC_nominal}) reduces the sample complexity while simultaneously reducing the number of decision variables in the optimization problem, thus resulting in a more computationally efficient scheme.

For the DDPC scheme in \eqref{eqn_DDPC}, note that the dimension of the decision variable $\alpha(t)\in\mathbb{R}^{T-L-n+1}$ increases as the number of offline data points $T$ increases (compare the lower bound for $T$ above). This implies that for increasing system dimension, number of inputs and prediction horizon length, the problem in \eqref{eqn_DDPC} becomes more expensive to solve. To address this issue, the authors of \cite{zhang2022} implement a pre-processing step of computing the singular value decomposition of the Hankel matrix of data $\mathscr{H}_{L+n}(w^{\textup{data}})$ in \eqref{eqn_DDPC}. Let such decomposition take the following form
\begin{equation}
	\mathscr{H}_{L+n}(w^{\textup{data}}) = \begin{bmatrix}
		U_1 & U_2
	\end{bmatrix}\begin{bmatrix}
	S_1 & 0\\ 0 & 0
\end{bmatrix}\begin{bmatrix}
V_1^\top\\ V_2^\top
\end{bmatrix},
\end{equation}
where $S_1$ is a diagonal matrix that contains the non-zero singular values of $\mathscr{H}_{L+n}(w^{\textup{data}})$, while $U_{i},V_i$, for $i=\{1,2\}$, are matrices of appropriate dimensions. As a consequence of the singular value decomposition, it holds that im$(U_1S_1)=\textup{im}(\mathscr{H}_{L+n}(w^{\textup{data}}))$. Therefore, the authors propose using $U_1S_1$ as a predictor in a predictive control scheme. Since $U_1S_1$ has $m(L+n)+n$ columns, the corresponding regressor vector (denoted $g\in\mathbb{R}^{m(L+n)+n}$) is of reduced dimension. However, one still requires the same number of data points needed for the DDPC (in \eqref{eqn_DDPC}) in order to build the matrix $\mathscr{H}_{L+n}(w^{\textup{data}})$ and compute its SVD, making this scheme not sample-efficient.

In the following section, we propose a data-driven predictive control scheme which is both more sample-efficient and more computationally efficient than existing schemes in the literature. This is done by implementing a few (algebraic) pre-processing steps on the collected data, which can be done prior to the online phase of the DDPC scheme.

\subsection{Nominal eDDPC scheme}\label{sec_eDDPC_nominal}
When comparing the results of Lemma~\ref{lemma_FL} and Theorem~\ref{thm_AFL}, we see that both provide us with a non-parametric representation of the finite-length behavior of a controllable LTI system. This is true since im$(\mathscr{H}_L(w^{\textup{data}}))=\textup{im}(P)=\mathscr{B}|_L$. However, the following are two important distinctions between the two results that will enable us to arrive at an efficient data-driven predictive control scheme.
\setlength{\textfloatsep}{1em}%
\begin{algorithm}[!t]
	\caption{Offline data pre-processing for eDDPC}\label{alg_preprocessing}
	\textbf{Input:} Measured trajectory $w^{\textup{data}}\in\mathscr{B}|_T$, satisfying rank$(\mathscr{H}_{d}(w^{\textup{data}}))=md+n$ for $d=\ell+1$, where $\mathscr{B}\in\mathscr{L}_{m,n,\ell}^{q}$.
	\begin{itemize}
		\item[1.] Compute a basis for the left kernel of $\mathscr{H}_d(w^{\textup{data}})$, i.e., $R_d\mathscr{H}_d(w^{\textup{data}})=0$.
		\item[2.] Use $R_d$ to build the matrix $\Gamma$ as in \eqref{Bmatrix}, but with $L+n-d$ shifts.
		\item[3.] Obtain $P=\textup{null}(\Gamma)$.
	\end{itemize}
	\textbf{Output:} Matrix $P$ such that im$(P)=\mathscr{B}|_{L+n}$.
\end{algorithm}%
\begin{itemize}
	\item[\textbf{D1}] Theorem~\ref{thm_AFL} requires at least $T\geq (m+1)(\ell+n+1)-1$ data points to satisfy the PE condition, whereas Lemma~\ref{lemma_FL} requires at least $T\geq (m+1)(L+n)-1$. If the minimum $T$ is chosen in both cases, then Theorem~\ref{thm_AFL} will always require $(m+1)(L-\ell-1)$ less samples, for any $L>\ell+1$.
	\item[\textbf{D2}] Given $\bar{w}\in\mathscr{B}|_L$, equation \eqref{GPE_FL} has infinitely many solutions for $\alpha\in\mathbb{R}^{T-L+1}$, whereas the corresponding $\beta\in\mathbb{R}^{mL+n}$ vector in \eqref{eqn_efficientFL} is \textit{unique for each trajectory}. This is because in order to fix a trajectory of an LTI system, one would only need $mL+n$ degrees of freedom that correspond to fixing the initial state ($n$) and specifying the input trajectory ($mL$). Notice that the dimension of $\beta$ is \textit{independent} of the number of collected data points, whereas the dimension of $\alpha$ increases with increasing $T$. In fact, even if the minimum $T$ was chosen for the results of Lemma~\ref{lemma_FL}, then the dimension of $\beta$ in Theorem~\ref{thm_AFL} would still be smaller than the dimension of $\alpha$ by $mn$.
\end{itemize}

We now exploit \textbf{D1} and \textbf{D2} to propose an efficient data-driven predictive control scheme (eDDPC). Recall that in order to arrive at the data-based representation in \eqref{eqn_efficientFL}, Theorem~\ref{thm_AFL} implements a few (algebraic) pre-processing steps on the collected data. In Algorithm~\ref{alg_preprocessing}, we summarize these steps which can be efficiently done offline. Since the predicted trajectories are of length $L+n$ (to account for the initialization at each time), Step 2 of Algorithm~\ref{alg_preprocessing} requires that we carry out $L+n-d$ shifts compared to $L-d$ shifts as in \eqref{Bmatrix}.

Once these (algebraic) pre-processing steps have been implemented, we obtain a matrix $P$ such that im$(P)=\mathscr{B}|_{L+n}$. Now, we construct the following minimization problem, which is solved at each time step $t$ and defines our efficient data-driven predictive control scheme (eDDPC)
\begin{equation}
	\begin{aligned}
		\min_{\beta(t),\bar w(t)}\quad 
		& \sum_{k=0}^{L-1}\norm*{\bar w_k(t) - w^s}_W^2\\
		\textrm{s.t.}\quad 
		& \bar w_{[-n,L-1]}(t) = P\beta(t)\\
		& \bar w_{[-n,-1]}(t) = w_{[t-n,t-1]}\\
		& \bar w_{[L-n,L-1]}(t) = w^s_n\\
		& \bar w_k(t) \in \mathbb{W},\quad \forall k \in \mathbb{Z}_{[0,L-1]}.
	\end{aligned}\label{eqn_eDDPC}
\end{equation}

Apart from using $P$ to predict future trajectories in place of the Hankel matrices, the two problems in \eqref{eqn_DDPC} and \eqref{eqn_eDDPC} have the same cost function and the same constraints. The proposed eDDPC scheme is summarized in Algorithm~\ref{alg_eDDPC}.

\begin{algorithm}[!t]
	\caption{eDDPC scheme}\label{alg_eDDPC}
	\textbf{Input:} Measured trajectory $w^{\textup{data}}\in\mathscr{B}|_T$, satisfying rank$(\mathscr{H}_{d}(w^{\textup{data}}))=md+n$ for $d=\ell+1$, where $\mathscr{B}\in\mathscr{L}_{m,n,\ell}^{q}$.\\
	\textbf{Offline phase:} Run Algorithm~\ref{alg_preprocessing} to obtain matrix $P$.\\
	\textbf{Online phase:}
	\begin{itemize}
		\item[1.] At time $t$, use past $n$ input-output measurements $w_{[t-n,t-1]}$ to solve \eqref{eqn_eDDPC}.
		\item[2.] Apply the input $u_t=\bar{u}^*_0(t)$ to the system.
		\item[3.] Set $t=t+1$ and return to 1.
	\end{itemize}
\end{algorithm}%

Since im$(P)=\textup{im}(\mathscr{H}_{L+n}(w^{\textup{data}}))$, the two optimization problems in \eqref{eqn_DDPC} and \eqref{eqn_eDDPC} are in fact equivalent and the resulting closed-loop trajectories are identical. The following lemma formalizes the equivalence between our proposed eDDPC scheme and the one from \cite{Berberich203}.
\begin{lemma}
	Given $w\in\mathscr{B}|_T$ where $\mathscr{B}\in\mathscr{L}_{m,n,\ell}^{q}$, let $\textup{im}(\mathscr{H}_{L+n}(w^{\textup{data}}))=\textup{im}(P)=\mathscr{B}|_{L+n}$ where $P$ is obtained by Algorithm~\ref{alg_preprocessing}. For a given initial trajectory $w_{[t-n,t-1]}$, if \eqref{eqn_DDPC} and \eqref{eqn_eDDPC} are initially feasible, then the two DDPC schemes result in the same closed-loop trajectories $u_t,y_t$ for~$t\geq0$.
\end{lemma}
\begin{proof}
	The result directly follows from the the fact that im$(P)=\textup{im}(\mathscr{H}_{L+n}(w^{\textup{data}}))$.
\end{proof}
\begin{remark}
	In \cite{Fiedler21} equivalence was established between the DeePC scheme of \cite{Coulson20} (which solves at each time $t$ a similar problem to \eqref{eqn_DDPC}, but without terminal constraints) and the subspace predictive control scheme (SPC) of \cite{Favoreel99}. A similar equivalence result can be established between SPC and eDDPC \eqref{eqn_eDDPC}. This is omitted from this paper for space reasons, but also follows from the fact that $\textup{im}(P)=\textup{im}(\mathscr{H}_{L+n}(w^{\textup{data}}))$.
\end{remark}
Another implication of the equivalence of the two optimization problems \eqref{eqn_DDPC} and \eqref{eqn_eDDPC} is that the proposed eDDPC scheme retains the same theoretical guarantees as the ones shown in \cite{Berberich203} for the DDPC scheme specified by \eqref{eqn_DDPC}.

In the next section, we analytically compare the number of data points and the number of decision variables needed for both schemes. Later we illustrate how the proposed eDDPC scheme outperforms the existing schemes in \cite{Berberich203, Dwyer23, zhang2022}.

%% file: sections/comparison.tex
\section{Comparison to existing schemes}\label{sec_eDDPC_comparison}
\begin{table*}[!t]
	\caption{Analytic comparison between eDDPC and existing DDPC schemes.}
	\begin{center}
		\begin{tabular}{ | c | c | c | c | c | }
			\hline 	
			& DDPC & sDDPC & SVD-DDPC & eDDPC\\ \hline
			$T\geq$ & $(m+1)(L+2n)-1$ & $(m+1)(2T_{\textup{ini}}+n)-1$ & $(m+1)(L+2n)-1$ & $(m+1)(\ell+n+1)-1$\\ \hline
			$\begin{matrix}
				\textup{dim(regressor)}
			\end{matrix}$ & $T-L-n+1$ & $n_s(T-2T_{\textup{ini}}+1)$ & $m(L+n)+n$ & $m(L+n)+n$\\
			\hline
		\end{tabular}
	\end{center}
	\label{table_analytic_comparison}
	\hrulefill
	\vspace{-1em}
\end{table*}
In this section, we analytically and numerically compare our proposed eDDPC scheme to the schemes reviewed in Section~\ref{sec_eDDPC_overview}. These include the DDPC scheme of \cite{Berberich203}, the segmented DDPC scheme of \cite{Dwyer23} and the minimum-dimension DDPC scheme of \cite{zhang2022}. Note that the authors of \cite{Dwyer23, zhang2022} do not enforce a terminal equality constraint. In the following, we implement such terminal constraints to their schemes and refer to them as sDDPC and SVD-DDPC, respectively. This ensures that the closed-loop trajectories are identical and that our comparisons are fair. As mentioned in the introduction, several other DDPC schemes have been proposed in the past few years. In this paper, we restrict the comparison to the schemes discussed above since (i) the closed-loop behaviors of all of them are equivalent and, hence, sample efficiency and computational efficiency are studied and compared fairly, and (ii) the schemes in \cite{Dwyer23} and \cite{zhang2022} explicitly address either sample efficiency \cite{Dwyer23} or computational efficiency \cite{zhang2022} as well.

Table~\ref{table_analytic_comparison} analytically compares the minimum number of required offline data points as well as the complexity of the corresponding optimization problems solved at each time $t$ (expressed in terms of the dimension of the regressor vector) for the following schemes: (i) the (DDPC) scheme of \cite{Berberich203}, (ii) the segmented DDPC (sDDPC) scheme of \cite{Dwyer23}, (iii) the minimum-dimension DDPC (SVD-DDPC) scheme of \cite{zhang2022} and finally (iv) our proposed (eDDPC) scheme \eqref{eqn_eDDPC}. It can be seen that the proposed eDDPC scheme outperforms the existing ones in terms of sample complexity (uses less offline data) and computational complexity (includes less decision variables).

Compared to the SVD-DDPC scheme, eDDPC uses the same number of decision variables but uses less data points. For the sDDPC scheme, the length of each segment must satisfy $T_{\textup{ini}}\geq\ell$. Here, we assumed that the prediction horizon is an integer multiple of $T_{\textup{ini}}$, i.e., $L=n_sT_{\textup{ini}}$ where $n_s$ is the number of segments. In this case, sDDPC and eDDPC schemes use the same number of data points only when $T_{\textup{ini}}=\ell=1$. Otherwise, eDDPC always uses less data points. As for the number of decision variables, eDDPC always uses less decision variables, even when $T$ is minimal.

Next, we carry out numerical simulations on random LTI systems to illustrate the efficiency of the eDDPC scheme. We consider systems of dimensions $n=\{4,6,8,10,12,14\}$ (with $p=m=n-2$). For each system order, we first generate 100 random systems (using Matlab's built-in \texttt{drss} command) and collect data by applying a PE input (sampled from a random a uniform distribution $\mathcal{U}(-1,1)^m$) with the minimum possible length $T$ as specified in Table~\ref{table_analytic_comparison}. Notice that for a fixed system order, number of inputs and number of outputs, the lag is not necessarily equal across all 100 systems. Therefore, for consistency we use a slightly longer data sequence for eDDPC than the minimum possible one specified in Table~\ref{table_analytic_comparison}. Specifically, since $\ell\leq n$, we collect $T=(m+1)(2n+1)-1$ data points for eDDPC. Furthermore, in order for all four schemes to have the same initialization $w_{[t-n,t-1]}$, we set $T_{\textup{ini}}=n$ for the sDDPC scheme. Notice that in this considered setting, eDDPC is still more sample-efficient than the other three schemes.

We then implement the four different DDPC schemes in a condensed formulation, i.e., the optimization problems of each scheme are reformulated such that the only decision variables are the regressors, e.g., $\alpha(t)$ in DDPC \eqref{eqn_DDPC} or $\beta(t)$ in eDDPC~\eqref{eqn_eDDPC}. The objective is to stabilize the origin. We use a prediction horizon $L=2n$, and sample the initial internal state from a uniform distribution $\mathcal{U}(-1,1)^n$ to later obtain the initialization trajectory $w_{[-n,-1]}$. The same quadratic stage costs (with $Q=R=I_{n-2}$) and the following input/output constraints were used in all schemes
\begin{equation}
	\mathbb{W}=\{w=\begin{bsmallmatrix}
		u\\y
	\end{bsmallmatrix}\in\mathbb{R}^{q}~|~ -5\leq w_i \leq 5,\quad\forall i\in\mathbb{Z}_{[1,q]}\}.
\end{equation}
This process is finally repeated and averaged for all 100 systems (per system dimension). The simulations were done on a standard Intel Core i7-10875H CPU @ 2.30GHz processor with 16 GB of RAM, using Matlab's quadprog with active-set algorithm and warm-start strategy. The corresponding MATLAB codes can be found at \cite{codes}.

Table~\ref{table_numerical_comparison} summarizes the results. It can be seen that our proposed eDDPC scheme outperforms the other schemes in terms of number of offline data points. We report both the average (avg) computation time per iteration over 100 systems as well as the maximum (max) computation time (for any MPC iteration out of all 100 systems). It can be seen that eDDPC and SVD-DDPC schemes have the fastest average computation time of all four schemes. Furthermore, their computation times are almost equal, which is expected since they have the same (minimum) number of decision variables among all four schemes. It can be also seen that the computation time of DDPC is of the same order of magnitude as eDDPC and SVD-DDPC. This is because we use the minimum number of data points and, hence, the difference in the number of decision variables is only $mn$ (cf. \textbf{D2} in Section~\ref{sec_eDDPC_nominal}). However, if more data points are used, the DDPC's average computation time increases as the dimension of $\alpha(t)$ grows. As for sDDPC, it should be noted that it uses less data points than DDPC and SVD-DDPC, however it has the slowest average computation times due to the increased number of decision variables caused by the segmentation (cf. \cite{Dwyer23}).
\begin{table*}[!t]
	\caption{Numerical simulations on 100 random systems of different complexities.}
	\vspace{-1em}
	\begin{center}
		\begingroup
		\setlength{\tabcolsep}{5pt}
		\begin{tabular}{ | c | cccc | cccc | cccc | cccc | }
			\hline
			& \multicolumn{4}{c|}{DDPC}& \multicolumn{4}{|c|}{sDDPC} & \multicolumn{4}{c|}{SVD-DDPC} & \multicolumn{4}{c|}{eDDPC}\\
			
			\hline
			States  & $\begin{matrix}
				\textup{avg}\\ [\mathrm{ms}]
			\end{matrix}$ & $\begin{matrix}
				\textup{max}\\ [\mathrm{ms}]
			\end{matrix}$ & $T$ & dim($\alpha$) & $\begin{matrix}
				\textup{avg}\\ [\mathrm{ms}]
			\end{matrix}$ & $\begin{matrix}
				\textup{max}\\ [\mathrm{ms}]
			\end{matrix}$ & $T$ & $\sum$dim($g_i$) & $\begin{matrix}
				\textup{avg}\\ [\mathrm{ms}]
			\end{matrix}$ & $\begin{matrix}
				\textup{max}\\ [\mathrm{ms}]
			\end{matrix}$ & $T$ & dim($g$) & $\begin{matrix}
				\textup{avg}\\ [\mathrm{ms}]
			\end{matrix}$ & $\begin{matrix}
				\textup{max}\\ [\mathrm{ms}]
			\end{matrix}$ & $T$ & dim($\beta$)\\
			\hline 
			$n=4$ & 1.19 & 6.98 & 47 & 36 & 1.68 & 3.94 & 35 & 56 & 1.06 & 4.22 & 47 & 28 & 1.04 & 2.94 & 26 & 28\\
			$n=6$ & 2.76 & 6.04 & 119 & 102 & 4.67 & 7.44 & 89 & 156 & 2.40 & 4.19 & 119 & 78 & 2.37 & 4.52 & 64 & 78\\  
			$n=8$ & 6.01 & 10.32 & 223 & 200 & 17.77 & 39.02 & 167 & 304 & 4.74 & 8.61 & 223 & 152 & 4.70 & 12.01 & 118 & 152\\
			$n=10$ & 20.07 & 40.55 & 359 & 330 & 66.98 & 131.23 & 269 & 500 & 14.63 & 48.00 & 359 & 250 & 14.77 & 48.30 & 188 & 250\\
			$n=12$ & 45.70 & 75.60 & 527 & 492 & 174.32 & 325.75 & 395 & 744 & 32.36 & 95.61 & 527 & 372 & 32.18 & 58.67 & 274 & 372\\
			$n=14$ & 97.41 & 161.50 & 727 & 686 & 448.11 & 695.54 & 545 & 1036 & 67.22 & 116.83 & 727 & 518 & 67.03 & 111.76 & 376 & 518\\
			\hline	
		\end{tabular}
		\endgroup
	\end{center}
	\label{table_numerical_comparison}
	\hrulefill
	\vspace{-1em}
\end{table*}

The results of Tables~\ref{table_analytic_comparison} and \ref{table_numerical_comparison} highlight the advantages of our proposed eDDPC approach, specifically in presence of scarce data and/or limited computational resources.

\begin{remark}\textit{Numerical considerations:}
	As the system complexity $(m,n,\ell)$ increases, some numerical problems were encountered when constructing the matrix $\Gamma$ in \eqref{Bmatrix}. Specifically, taking the first $p$ rows of the matrix $R_d$ can sometimes result in a matrix $\Gamma$ with a very large condition number, which might lead to errors in computing $P=\textup{null}(\Gamma)$. One explanation is that the kernel representation specified by the rows of $R_d$ is not necessarily minimal. In a minimal kernel representation, the number of rows of $R_d$ is equal to the number of outputs $p$ (cf. \cite{Willems86}), which is the number of rows that are shifted in~\eqref{Bmatrix}. Since obtaining the minimal kernel representation for multivariable systems is a difficult problem (cf. \cite{Markovsky22}), we devised a (heuristic) combinatorial method to obtain a well-conditioned matrix $\Gamma$. The idea is to try different combinations of $p$ rows of $R_d$ and use them to construct the shifts in $\Gamma$ and later check if the condition number is smaller than a specified threshold. Note that these steps are done offline and do not increase the computational burden of the eDDPC scheme.
\end{remark}

%% file: sections/conclusions.tex
\section{Conclusions}\label{sec_conclusions}
In this paper, we proposed a data-driven predictive control scheme which is both more sample-efficient (uses less offline data) and computationally efficient (uses less decision variables) than existing data-driven predictive control schemes in the literature. We analytically and numerically showed that the proposed scheme outperforms existing schemes from the literature. Future work will focus on providing theoretical guarantees for recursive feasibility and stability of a robust eDDPC scheme in presence of noise.